# Hidden Vibronic and Excitonic Structure and Vibronic Coherence Transfer in the Bacterial Reaction Center

Veronica R. Policht[1†], Andrew Niedringhaus[1], Rhiannon Willow[1], Philip D. Laible[2], David F. Bocian[3], Christine Kirmaier[4], Dewey Holten[4], Tomáš Mančal[5] and Jennifer P. Ogilvie[1]*

[1]Department of Physics, University of Michigan, 450 Church St, Ann Arbor MI 48109
[2]Biosciences Division, Argonne National Laboratory, Argonne, IL 60439, USA
[3]Department of Chemistry, University of California, Riverside, CA 92521, USA
[4]Department of Chemistry, Washington University, St. Louis, MO 63130, USA
[5]Faculty of Mathematics and Physics, Charles University, Ke Karlovu 5,
 CZ-12116 Prague 2, Czech Republic
[†]Present address: Politecnico di Milano, Department of Physics, 20133 Milan, Italy
*e-mail: jogilvie@umich.edu

**We report two-dimensional electronic spectroscopy (2DES) experiments on the bacterial reaction center (BRC) from purple bacteria, revealing hidden excitonic and vibronic structure. Through analysis of the coherent dynamics of the BRC we resolve specific coherent signatures that allow us to definitively assign the upper exciton energy of the "special pair." This assignment is supported by simulations of coherent dynamics of a reduced excitonic model of the BRC. The simulations also identify nonsecular vibronic coherence transfer processes neglected in standard models of photosynthetic energy transfer and charge separation. In addition, the coherent dynamics reveal multiple quasi-resonances between key intramolecular pigment vibrations and excited state energy gaps in the BRC. The functional significance of such electronic-vibrational resonances for photosynthetic energy transfer and charge separation remains an open question.**

In the primary steps of photosynthesis, light-harvesting antenna structures gather solar energy and transfer it to reaction centers (RCs) for processing[1]. Compared to the colorful array of photosynthetic antenna architectures that exist in nature, the RC structures are more widely conserved. The bacterial RC (BRC) from purple bacteria[2] features a pseudo two-fold-symmetric hexameric core of pigments (Supplementary Fig. 1) that converts excitation energy to a stable charge-separated state. The near unity quantum efficiency of the charge separation process is remarkable[2,3], as is the high specificity with which charge separation occurs along the "A" branch of the BRC structure[2,4]. In contrast to the Photosystem I and Photosystem II RCs in oxygenic photosynthesis, the BRC possesses a strongly coupled "special pair" of bacteriochlorophyll a (BChl), yielding greater spectral separation between the absorption features in the $Q_y$ region (Fig. 1b) making the BRC a simpler system for resolving the ultrafast processes of energy transfer and charge separation[5]. Understanding the design principles of photosynthetic systems may open new avenues for improving artificial solar harvesting devices and has helped motivate the development of new subfields of spectroscopy and theoretical approaches to describing the nonequilibrium photosynthetic process.

The combination of ultrafast timescales (~fs - ps) on which photosynthetic systems perform energy transfer and charge separation, and their broad absorptions arising from multiple coupled pigments in carefully-tuned dielectric environments present significant challenges to uncovering their structure-function relationship. Two-dimensional electronic spectroscopy (2DES) can address many of these challenges and has become a powerful tool for studying photosynthetic systems[6–9]. 2DES studies of the Fenna-Matthews-Olson (FMO) complex demonstrated its ability to uncover energy transfer pathways[6] and revealed long-lived (~ps) coherent oscillations[10,11]. Similar coherent processes have been reported in other photosynthetic systems[8]. Ultrafast pump probe studies of the BRC by Vos and Martin[12,13] in the early '90s provided the first observations of coherent dynamics in photosynthetic systems. These studies proposed that the coherences arose from vibrational wave packet motion on the excited electronic state delocalized across the special pair and surrounding protein matrix[13] and might facilitate electron transfer[14]. The initial 2DES experiments to observe long-lived coherences in FMO[10] proposed superpositions of delocalized electronic excited states as the origin of the coherence. It was later noted that the frequencies of the observed coherences matched pigment vibrational modes as well as excitonic energy gaps, raising a question about the role of vibrations in explaining both the observations of coherence and the high efficiency of energy[15–18] and electron transfer[19]. Since that time, theoretical and experimental work has aimed to establish how coherence between states of various physical origin (electronic, vibrational, and vibronic) is manifest in 2DES data[7,16,20–24]. It is now understood that vibrational degrees of freedom play a dominant role in the coherent dynamics observed in 2DES studies of photosynthetic antennas[16,21,25] and RCs[26–31]. What is less understood is the prevalence of electronic-vibrational resonances and their possible functional relevance[14–16,30–32].

Here we report coherence signatures from broadband 2DES studies of the neutral BRC as it undergoes energy transfer and charge separation, and compare these signatures to those obtained from monomeric BChl[33] as a control for purely vibrational coherence. The coherence signatures allow us to definitively confirm the position of the weak upper exciton state of the special pair that we recently proposed based on kinetic analysis of our 2DES data[34]. The coherences also reveal numerous quasi-electronic-vibrational resonances, leading to mixed vibronic transitions in the BRC that are only allowed by excitonic delocalization involving vibrationally excited states. Theoretical modeling confirms our assignment of the upper special pair exciton and suggests that certain coherence signatures are to a significant extent due to nonsecular coherence transfer among the BRC vibronic states during rapid downhill energy transfer and relaxation preceding charge separation. These nonsecular processes are frequently neglected in models of photosynthetic energy transfer and charge separation, including so-called "armchair" coherent modes which evolve in the ground state of one site and provide a window for observing excited state dynamics elsewhere in the BRC via vibronic coupling. Our work motivates further theoretical modeling that can assess the importance of nonsecular processes as well as quasi electronic-vibrational resonances for photosynthetic function.

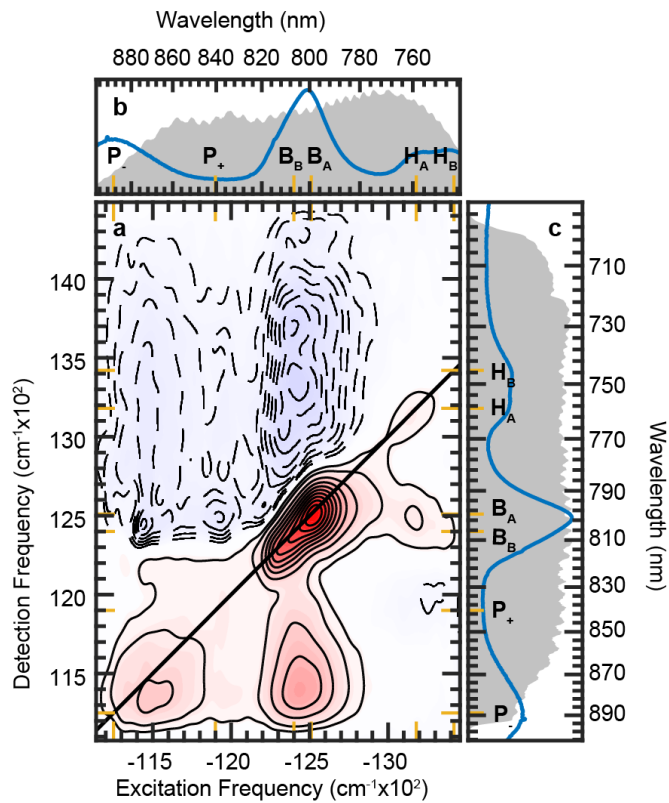

**Figure 1** (**a**) Absorptive (real, phased) 2D spectrum of the M250V BRC at a waiting time of 200 fs at 77 K. Along the top (**b**) and right (**c**) axes are shown the pump and probe spectra (shaded gray) used in the experiments, as well as the 77K BRC linear absorption spectra (blue). Gold lines along the axes indicate the positions of the excitonic states of the BRC[34].

**Results and Discussion**

*Two-dimensional electronic spectroscopy of the BRC*

We performed 2DES experiments on the W(M250)V BRC mutant of *Rhodobacter capsulatus* (Fig. 1a), which lacks the A side quinone and performs charge separation to the $P^+H_A^-$ state[35]. Figure 1a shows a typical absorptive 2DES spectrum of the BRC at a waiting time ($t_2$) of 200 fs and at 77 K. The pump and probe laser spectra are plotted along the respective excitation and detection axes in addition to the 77 K linear absorption spectrum of the BRC (Fig. 1b & c). Structure-based models of the BRC explain the three main peaks in the $Q_y$ linear absorption spectrum as arising in increasing energy from the lower (P-) exciton of the special pair, the combined BChl $B_A$ and $B_B$ pigments (B), and the bacteriopheophytins $H_A$ and $H_B$ (H) (see Supplementary Fig 1 for BRC structure). The electronic coupling between the special pair pigments ($P_A$, $P_B$) is sufficiently strong that the two lowest excited states largely consist of linear combinations of excitations of $P_A$ and $P_B$ pigments and are denoted the upper ($P_+$) and lower ($P_-$) exciton states. While the other electronic excited states are primarily localized on the B and H pigments, throughout this paper we use the term exciton to refer to any eigenstate of the coupled BRC. The location of $P_+$ has been historically difficult to assign given its weak oscillator strength and proximity to the strong ~800 nm absorption of the B pigments[36]. At 200 fs, cross-peaks are readily resolved in the 2DES spectrum between H and B, as well as B and P-, reflecting the excitonic coupling and rapid downhill energy transfer and equilibration that has already occurred (Fig. 2). The positions of the excitonic states are marked as solid gold lines in Figure 1 and were determined by multi-excitation 2D global analysis of the 2DES data as described by Niedringhaus et al[34]. As a control for purely vibrational coherence effects, we also present 2DES data of monomeric BChl (Supplementary Fig. 2)[33].

2DES spectroscopic signals can be analyzed in terms of so-called *Liouville pathways*, which describe the photo-induced time evolution of specific elements of the system's density matrix as represented in the eigenstate (exciton) basis. Intramolecular processes associated with various spectroscopic signals can be revealed by depicting the Liouville pathways using *double-sided Feynman diagrams* (Supplementary Fig. 3). A complete collection of the Liouville pathways and their associated signals provides complete information about the time evolution of the system's photo-induced state. Although disentangling such a complete set of all the required Liouville pathways from experimental data is usually not feasible, coherence between two eigenstates produces beating signals at the eigenstate difference frequency that can be selectively revealed by Fourier transform in the so-called waiting time $t_2$. Identifying spectral features associated with the Liouville pathways involving coherences provides unique insight into the photo-induced dynamics of the system. The dynamics of photosynthetic systems have typically been assumed to obey the so-called *secular approximation*; namely that the evolutions of the populations and coherences are independent of each other, and coherence transfer processes are neglected. In the secular approximation Liouville pathways involving excited or ground state coherence during the time interval $t_2$ can only decay, and cannot drive other coherences or populations, nor can they be fed from other coherences or populations.

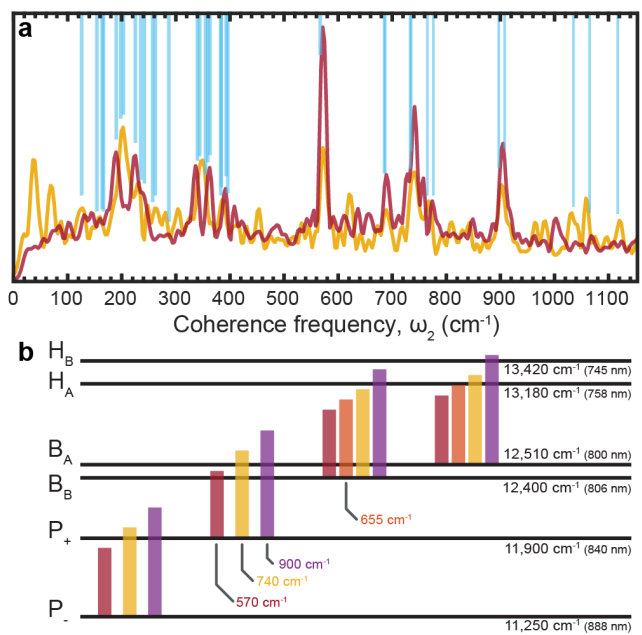

**Figure 2(a)** Frobenius spectra for the W(M250)V BRC (red) and bacteriochlorophyll a (gold). Blue lines indicate vibrational modes of BChl and Bacteriopheophytin a[37–39] that lie within the experimental resolution (9.7 cm$^{-1}$) of the most prominent peaks in the Frobenius spectra. Peak prominence is determined by ≥15% of the Frobenius spectrum maximum above the background noise level. It is worth noting that the BRC Frobenius spectrum is largely dominated by B-band contributions. Several less-prominent modes not highlighted here are also in good agreement with vibrational modes such as the $\omega_2 = 656$ cm$^{-1}$ mode (Supplementary Fig. 4). **(b)** The excitonic levels with energies taken from Niedringhaus et al.[34] demonstrates how several prominent coherence modes are in quasi-resonance with the energetic gap between excitonic levels.

To reveal the dominant frequencies of the coherences observed throughout the entire BRC 2D spectrum during the waiting time $t_2$ we take the Frobenius Norm of the 2D spectra Fourier transformed with respect to $t_2$ (Fig. 2a, red curve). For comparison we also show the Frobenius spectrum of BChl in isopropanol (gold curve)[33]. Above the Frobenius spectra we indicate vibrational modes reported in resonance Raman experiments of BChl[37–39] and bacteriopheophytin[39] which show good agreement with our measurements. Figure 2a clearly shows many prominent modes in the BRC with peak frequencies corresponding well with previous studies of coherence in the BRC[26–29]. The BChl spectrum shows many prominent modes (Fig. 2a) which have been previously attributed to purely vibrational coherences and are well described by a displaced oscillator model[33,40–42]. There is very high agreement of the peak locations in the monomer and BRC Frobenius spectra. By comparing the BRC peak frequencies with the Frobenius spectrum of monomeric BChl and vibrational spectroscopy literature for BChl monomers[37–39], we can assign many of the BRC peaks as largely vibrational in character. We note a large discrepancy in the low frequency region of the Frobenius spectra where the single peaks in the monomer spectrum at ~200 cm$^{-1}$ and ~340 cm$^{-1}$ appear split into two neighboring peaks in the BRC spectrum. This splitting of the low-frequency peaks is consistent with resonance Raman experiments which selectively excited the P- and B bands[37,43].

Figure 2b shows the excitonic states of the BRC derived from our fitting of the 2DES data[34] with superimposed dominant coherent modes from the Frobenius spectrum of Fig 2a drawn with height proportional to their energy and to scale with respect to the excitonic energy gaps. It is evident from Fig. 2b that there are numerous coherent modes that are quasi-resonant with excitonic energy gaps in the BRC. This suggests the existence of numerous quasi-electronic-vibrational resonances that can lead to vibronic delocalization and modify the allowedness of otherwise forbidden transitions. Such resonances have been reported to potentially enhance energy transfer in various antenna complexes[15–17] and charge separation in the photosystem II RC[14,30,31].

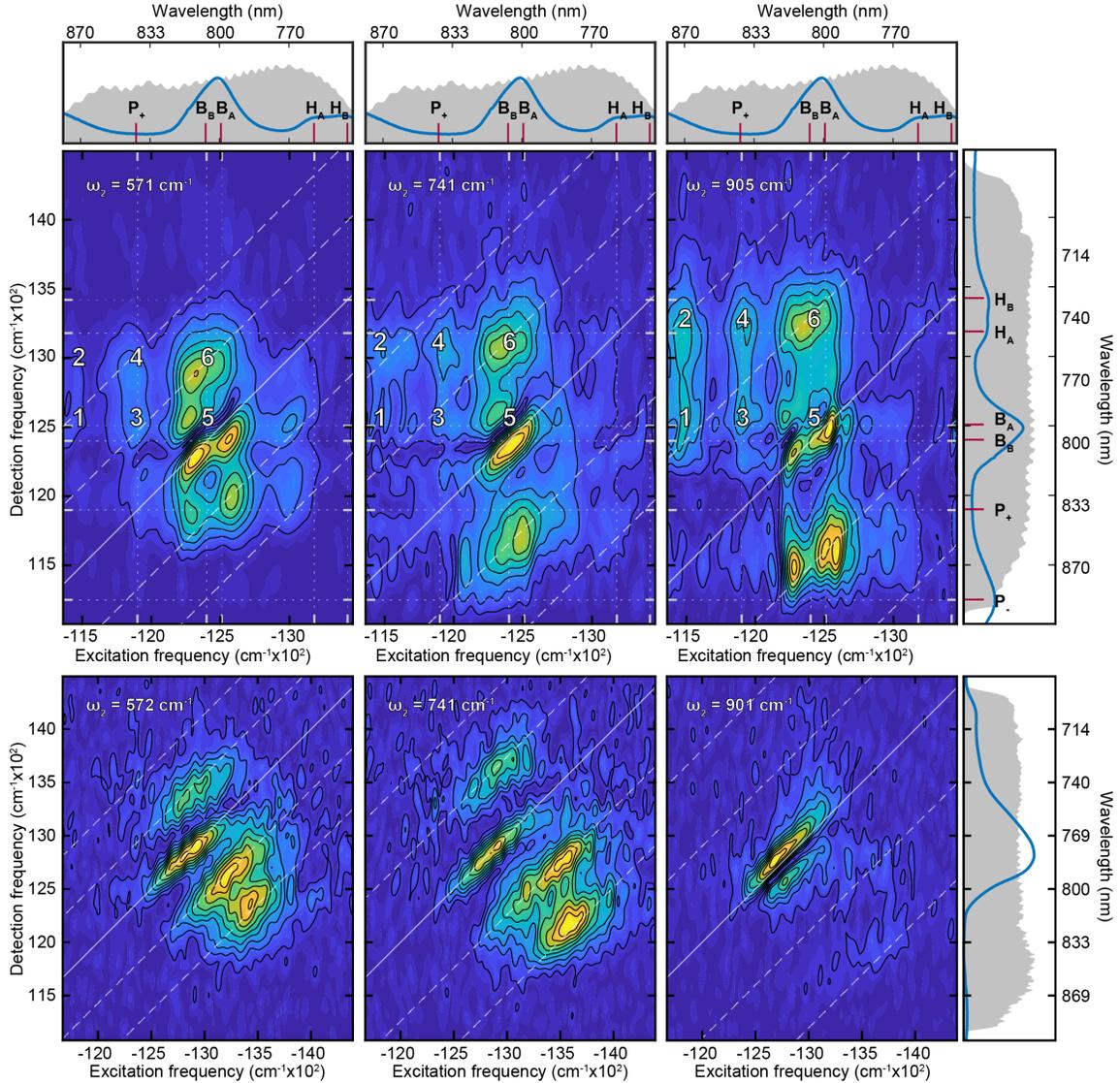

**Figure 3** Coherence amplitude maps of the BRC (top row) and monomeric BChl (bottom row) reveal the distribution of the observed coherent dynamics throughout the 2D spectra. Maps are derived from the real rephasing 2D spectra at $\omega_2$ values of 572 cm$^{-1}$, 741 cm$^{-1}$ and 905 cm$^{-1}$ (columns left to right). For each map, the diagonal ($\omega_{exc} = \omega_{det}$) is indicated by a solid diagonal line and parallel dashed white lines are offset from the diagonal by $\pm\omega_2$ and $-2\omega_2$ to aid in their interpretation. (Top row) BRC excitonic energies taken from Niedringhaus, et.al.[34] are indicated by white dotted vertical and horizontal lines. The 77K linear absorption spectrum and probe spectra are shown for easy reference for both BRC and BChl; we additionally show the pump spectrum for the BRC maps. (Top row) Signatures labeled 1-6 show evidence for vibronic coupling between the $P_+$, $P_-$ and $B_A$ states and are discussed in more detail in the text.

To better characterize the coherences, we examine the 2D amplitude distribution at individual coherence frequencies (Fig. 3). These coherence maps are obtained by plotting, at a given $\omega_2$, the absolute magnitude of the FFT of the oscillatory residual (after subtraction of the population kinetics) to each ($\omega_{exc},\omega_{det}$) point of the real rephasing 2D spectrum[22,25]. Further details are provided in the Supplemental Information, along with additional low frequency coherence maps of the real rephasing signals (Supplementary Fig. 4) and coherence maps of the complex rephasing signals showing separated positive and negative frequency components for the BRC (Supplementary Fig. 5) and BChl (Supplementary Fig. 6). By displaying the excitation and detection frequency dependence of the coherence amplitude we can more easily assign specific peaks to particular Liouville pathways characterizing the nonlinear response of the system (see Supplementary Fig. 8 & 9).

**Excitonic and Vibronic Structure in the BRC**

Differences between the BRC and monomer data that arise from excitonic and vibronic structure are evident in coherence maps of the higher frequency modes, where signals originating from distinct Liouville pathways can be identified. Figure 3 shows coherence maps of the BRC for the dominant modes in the Frobenius spectrum within quasi-resonance with excitonic energy gaps. Several of the coherence maps of the BRC show peaks centered around B (including peaks labeled 5 & 6) that are analogous to those observed in the BChl monomer maps[33] and can be explained within a simple displaced oscillator (DO) model (see Supplementary Fig. 8 for the assignment of these peaks).

The peaks labeled 1-4 in Figure 3 have no assignment within the standard displaced oscillator (Supplementary Fig. 8), electronic dimer or vibronic dimer models that neglect the doubly-excited excitonic state (Supplementary Fig. 9). Peaks 1 and 2 arise at an excitation energy of the lower exciton of the special pair ($P_-$) while we attribute the excitation energy of peaks 3 and 4 to the location of the upper exciton ($P_+$)[34]. To understand the origin of these spectroscopic signatures we constructed a Frenkel exciton model with a strongly coupled special pair homodimer and a single BChl pigment (Fig. 4a). While the site energy of the single B pigment is consistent with $B_A$, the simplicity of the model precludes a definitive assignment. Within our model the exciton energies and timescales of energy transfer and relaxation among the excitonic states are taken from the kinetic analysis of our 2DES data[34]. In addition, we include single vibrational modes with frequencies matching coherence frequencies in Figure 3. Details of the Frenkel exciton theory as well as links between the Liouville pathways and the 2DES spectral features can be found in the SI.

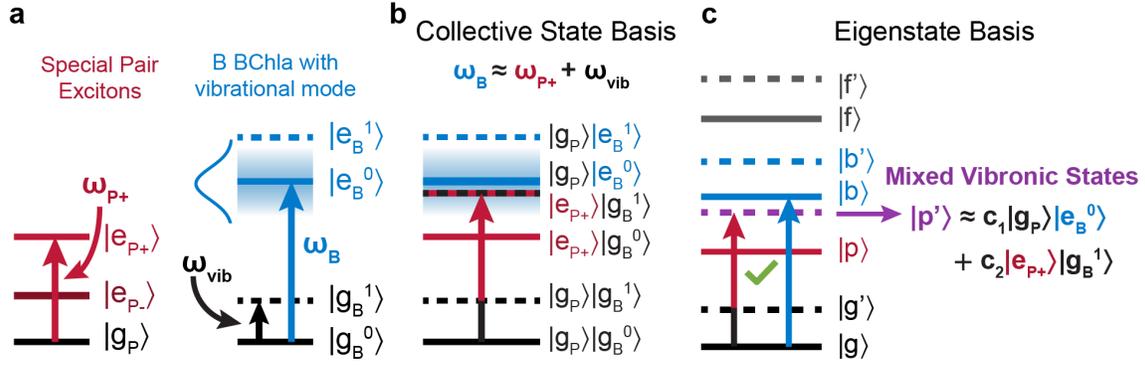

**Figure 4 (a)** The site basis representation of the reduced Frenkel exciton model where the strongly coupled special pair molecules are represented by their delocalized excitonic states ($|e_{P+}\rangle$, $|e_{P-}\rangle$) formed by electronic coupling between the special pair molecules, $J_{PP} = 325$ cm$^{-1}$ (STab. 2). The nearest-neighbor accessory BChl ($|e_B\rangle$) additionally has one excited vibrational state on the ground ($|g_B^1\rangle$) and excited ($|e_B^1\rangle$) electronic states. The inhomogeneously broadened excited electronic state of B ($|e_B^0\rangle$) is represented with a Gaussian distribution of 100 cm$^{-1}$ width (blue gradient). **(b)** The collective state basis represents the same states as in part **(a)** demonstrating the near resonance of two key transitions: $|g_P\rangle|e_B^0\rangle$ and $|e_{P+}\rangle|g_B^1\rangle$. **(c)** The eigenstate basis includes electronic coupling, $J_{PB} = 100$ cm$^{-1}$ (STab. 2), between the P and B sites and results in mixing between the two near-resonant transitions $|g_P\rangle|e_B^0\rangle$ and $|e_{P+}\rangle|g_B^1\rangle$. The electronic mixing results enhances the allowedness of state $|p'\rangle$, which involves simultaneous excitation of the special pair excited state and the ground state coherence of B ($|e_{P+}\rangle|g_B^1\rangle$). Eigenstate $|p'\rangle$ is the key state giving rise to signatures 3 & 4 and the corresponding lower exciton special pair state in signatures 1 & 2 (Fig. 3). Parts **(b)** and **(c)** focus on the P upper exciton for visual simplicity; a more complete eigenstate diagram is presented in Fig. 5.

We find that the characteristic peaks 1-4 in the experimental spectra (Fig. 3) can only be explained when the vibrational mode resides on the B pigment (Fig. 4a) rather than on one of the special pair pigments (Supplementary Fig. 28-31). This is unexpected, as the positions of the peaks suggest that the first excitation of the system involves P$_+$ or P$_-$, and appears unrelated to the formation of the vibrational coherence on B. Furthermore, we find that the 1-4 peak pattern is observed only when the energy of the vibrational quantum is in quasi-resonance with the energy gaps between two electronic eigenstates (see Supplementary movies). In quasi-resonance, vibrational states can borrow oscillator strength from the electronic transitions by forming mixed vibrational-electronic (vibronic) eigenstates, yet this mixing is weak due to energy gap mismatch (Fig. 4c), such that the new energy gaps (between eigenstates) are not significantly changed by the level repulsion. It was previously found that the resonance condition for enhancement of vibrational coherence in 2DES is rather broad[21] and does not require an exact resonance between electronic and vibrational energy gaps. Under the quasi-resonance conditions, the observed oscillatory frequency is almost identical to the vibrational frequency of the original intramolecular mode and the eigenstates of the RC largely keep their original electronic character. However, some previously forbidden states, notably those containing vibrational excitation in the electronic ground state of molecules, acquire non-zero transition dipole moment (Fig. 4b & c). At the same time, the condition of quasi-resonance allows vibrational modes with a broad range of frequencies to have enhanced spectroscopic signatures. As a result, different frequencies produce similar patterns in 2DES, consistent with our observations in Fig. 3.

Based on the simulations, we are able to assign a set of Liouville pathways to each peak labeled 1-6 in the experimental coherence maps (see Fig 3). An extensive list of associated Liouville pathways can be found in the SI. We highlight four types of coherence signatures of particular interest, some of which are nonsecular. Type i): decaying vibrational coherence residing in the ground or excited state of the B molecule. This is a secular process, as it involves only coherence decay and is therefore allowed under the secular approximation. Type ii): vibronic coherence shift from the excited state of B to its ground state accompanied by electronic energy transfer from the B molecule to the $P_-$ or $P_+$ states. This nonsecular process was previously identified by Paleček et al[27] and named Energy Transfer-Induced Coherence Shift (ETICS). Type iii): electronic to vibrational coherence transfer, in which an initial electronic coherence, such as the one between electronic excited states of the B molecule and the P. state of the SP is transferred into a coherence characterized by a population of an electronic state (e.g. P$_-$) with a vibrational coherence in the ground state of the B molecule. These processes can also formally be regarded as a manifestation of ETICS (see Fig 5c). Type iv): we refer to this novel type of nonsecular process as "armchair" coherence, where the ground-state coherence acts as a spectator of the excited electronic dynamics occurring elsewhere in the system. Its defining characteristic is that vibrational coherence in the ground state of B is maintained throughout the process. An example is given in Fig 5c, where armchair coherence enables identification of P$_+$ during energy relaxation from P$_+$ to P$_-$.

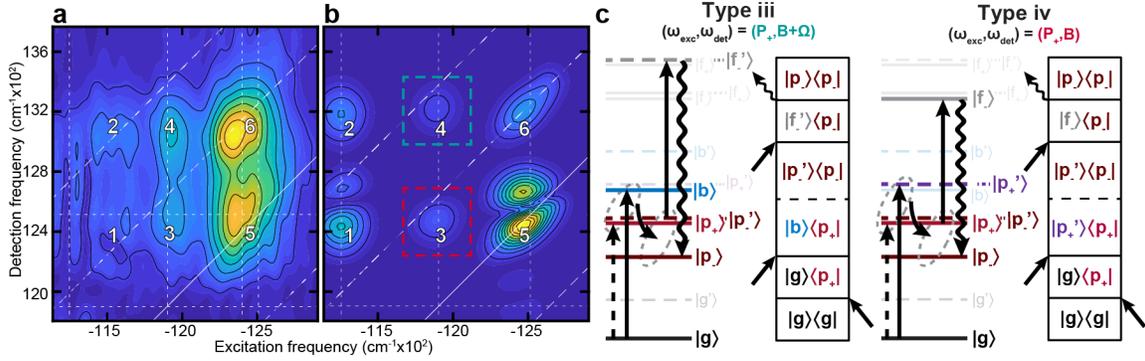

**Figure 5** Experimental (**a**) and simulated (**b**) complex rephasing coherence maps for $\omega_2 =$ +740 cm$^{-1}$ featuring the prominent signatures (labeled 1-6) of vibronic coherence between special pair and accessory BChl sites. (**c**) Dominant light-matter interaction pathways for signatures 3 & 4 identified from simulations and represented both as light-matter Jablonski Diagrams (left) and Double-Sided Feynman Diagrams (right). These pathways are examples of the general classes of coherence types iii and iv described in the text. Both type iii and iv coherence signals are dominant in signatures 3 & 4, where the third light-matter interaction (probe) determines the pathway's detection frequency.

Our simulations (Fig. 5) confirm our assignment of peaks 1 and 2 to the Liouville pathways starting with the lower exciton state P$_-$ and peaks 3 and 4 to the upper exciton state P$_+$ (Fig. 3). The process of vibrational coherence decay denoted as Type i) is the leading contributor in terms of amplitude to peaks 1 and 2 of the calculated spectra (Supplementary Fig. 16, 17, 19, & 20). The striking feature of this coherence is the fact that although the process occurs nominally in the collective excited state of the BRC, it involves a vibrational mode excited in the electronic ground state of the B molecule. As is well-known, excitonic states involve a single molecular excitation (possibly delocalized over several molecules) while the remaining molecules of the aggregate remain unexcited. Due to resonance interaction between the molecules, some allowed excitonic states contain vibrational excitations on the electronic ground state of participating molecules (see Fig. 4 and the SI for detailed information on the composition of vibronic eigenstates of the model BRC). To excite from the electronic (and vibrational) ground-state of the aggregate to what is effectively a ground state vibration, the present mechanism requires only a single interaction with the laser pulse. This makes the mechanism of vibronic enhancement at work here different from the one discussed by Tiwari et al.[16], where the observed oscillations are part of the ground-state bleach (GBS) signal wherein the vibrational coherence in the electronic ground state is excited after two interactions of the field with the system. Type i) coherence is more closely related to the excited state vibronic coherence described by Christensson et al.[21], with the distinction that the oscillating signal measured in the present work corresponds to excited state absorption rather than stimulated emission.

Peaks 3 and 4 (Fig. 3) are dominated by Liouville pathways involving energy transfer from state $P_+$ to $P_-$ in a process of Type iv) (Fig. 5c). Here we first excite the state $P_+$ and the vibrational coherence on the B molecule. During $t_2$, population of $P_+$ transfers to $P_-$ while the vibrational coherence in the ground state of the B molecule remains largely unaffected. However, its presence allows us to observe the energy relaxation process in the coherence map. Another dominant pathway contributing to peaks 3 and 4 is a type of ETICS process (Type iii)), an example of which is given in Fig. 5c. A detailed account of the Liouville pathways identified in our simulations as leading contributions to signals 1-6 can be found in the SI.

We note the utility of "armchair" coherence (Type iv) in clearly revealing elusive states such as P₊. This state is weakly present in the real absorptive 2DES data in Fig. 1a and was assigned to an energy of 11,900 cm$^{-1}$ after extensive kinetic analysis[34]. In contrast, armchair coherence renders this state easy to detect. The ability of 2DES to reveal signatures of dark states through coherence map analysis has been demonstrated previously[44]. The assignment of P₊ is notable given the difficulty of determining the location of P₊ in previous studies of BRCs. Experimental studies of *R. sphaeroides* locate P₊ between 12,225 – 12,642 cm$^{-1}$ (818 – 791 nm)[45–47] at 1.5-10 K and at 12,121 cm$^{-1}$ (825 nm) at room temperature[48], while a theoretical model predicts P₊ at 12,346 cm$^{-1}$ (810 nm) and 12,285 cm$^{-1}$ (814 nm) at 77 K and room temperature, respectively[36]. The site energies and couplings of the special pair molecules also depend on experimental factors including temperature and solvent composition[4]. The discrepancies between our assignment of 11,900 cm$^{-1}$ (840 nm) to P₊ in the *R. capsulatus* BRC and BRCs in other works may also be a result of differences between bacterial species studied.

While the possibility of coherence transfer between excitonic states in photosynthetic systems has been discussed and it has been proposed that coherence transfer contributions to 2DES signals can be significant in strongly-coupled systems[49], most theories of photosynthetic energy transfer have focused on population transfer, ignoring coherence transfer processes via the secular approximation. Vibrational coherence transfer has been previously reported in 2D IR experiments[50], 2DES studies of silver nanoclusters[52] and transient absorption experiments probing passage through conical intersections[51]. The spectroscopic signatures we observe in the BRC suggest the possible importance of electronic-vibrational resonances and coherence transfer processes involving vibrations on B. These vibrations may play a functional role in assisting the transfer of electronic excitation energy. Such a role for electronic-vibrational resonances has been proposed in other photosynthetic systems[15,16], as has their possible importance for downhill energy transfer from H to B in the BRC[27]. Similarly, there has been ongoing debate about the importance of electronic-vibrational resonances for photosynthetic charge separation[14,32].

In conclusion, we report previously hidden excitonic and vibronic structure in the BRC, revealed through analysis of coherent 2DES signals. We confirm the assignment of the elusive special pair upper exciton state and find numerous quasi-electronic-vibrational resonances in the BRC. Through a comparison of the monomer BChl and BRC 2DES coherence maps along with simulations of a reduced BRC model we identify nonsecular coherence transfer processes involving the special pair and B pigments, in which vibrations on B play a prominent role. The possible functional importance of such processes, as well as quasi-electronic-vibrational resonances to photosynthetic energy transfer and charge separation merit further theoretical and experimental examination.

**Methods Summary**

All experimental data presented here was collected using a hybrid pump-probe, background free 2DES setup which utilizes an acousto-optic programmable dispersive filter (AOPDF) pulse shaper (Dazzler, Fastlite) to simultaneously collect rephasing, nonrephasing, and transient grating signal[53]. Interpulse delay between the first and second pump pulses were scanned $0 < t_1 < 390$ fs in 10 fs steps; waiting time between second pump and the probe, $t_2$, was scanned in 10 fs steps from $-50 < t_2 < 3500$ fs. Data was truncated at $80 < t_2 < 3500$ fs prior to coherence analysis to eliminate contributions by coherent transients, resulting in a coherence frequency resolution of 9.8 cm$^{-1}$ and maximum resolvable coherence frequency of 1668 cm$^{-1}$. Light pulses were generated using a home-built optical parametric amplifier seeded by a 500 Hz, 40 fs, Tai:Sapph regen (SpitFire, Spectra Physics) centered at 800 nm. Both experiments used pumps, probe and local oscillator from the same degenerate OPA (DOPA)[54] (spectra shown in Fig.1a&b and Supplementary Fig 3a&b for the BRC and BChl experiments, respectively). The pump used in the BRC (BChl) experiment was compressed to 14.5 (14.3) fs with an AOPDF and the probe to 10.3 (10) fs using chirped mirrors and a SLM pulse shaper (femtoJock, Biophotonics).

All experiments shown here were conducted at 77 K in a LN$_2$ cryostat (Microstat, Oxford Instruments) with a 380 μm path length cell. Samples were prepared to have an average OD ≈ 0.3, though the BRCs were prepared with higher peak OD in order to better resolve the lower dipole-strength special pair. BChl samples were prepared from powder purchased from Sigma Aldrich and were dissolved in isopropanol, which was degassed using N$_2$ gas before sample preparation. The sample was handled and loaded under N$_2$ atmosphere in a glove box.

W(M250)V BRC mutants were isolated and kept in pH 7.8 10 mM Tris with 0.1% Deriphat buffer. Prior to concentration, BRC samples were treated with 40 mM terbutryn in order to remove or inactivate quinones and 400 mM sodium ascorbate to reduce the special pair between laser shots. Following concentration, BRCs in buffer were mixed with glycerol (1:1 v/v) to ensure good quality glass when frozen.

**Acknowledgments**

V. Policht, A. Niedringhaus and J. P. Ogilvie gratefully acknowledge the National Science Foundation (grant # PHY-0748470 and #PHY-1607570) for support of their salaries. J.P.O. also acknowledges support from the Office of Basic Energy Sciences, the US Department of Energy under grant number DE-SC0016384.

C. Kirmaier, D. Holten and P. D. Laible acknowledge funding from U. S. Department of Energy, Office of Basic Energy Sciences under grant DE-CD0002036 (C.K. and D.H.) and associated Field Work Proposal (P.D.L.). T. Mančal acknowledges the support from Charles University Research Center Program UNCE/SCI/010 and the Czech Science Foundation (GAČR) grant no. 20-011595. We thank Kevin Kubarych and Eitan Geva for helpful discussions.

Correspondence should be directed to J. P. Ogilvie (jogilvie@umich.edu).

Author Contributions

V. P., A. N and J. P. O. conceived of and designed the experiments: V. P. and A. N. performed the experiments: V. P. analyzed the data: P.D.L, C. K. and D. H. provided samples: T. M. performed the simulations: all of the authors discussed the data and its significance: V. P. and J. P. O. wrote the paper with input from all of the authors.

Competing Interests

The authors declare no competing interests.